\documentstyle[twocolumn,aps,eqsecnum]{revtex}
\begin{document}

\thispagestyle{empty}
\setcounter{page}{1}

\title{{\small\begin{flushright} CERN-TH/98-291
\end{flushright}}
HARD THERMAL LOOPS IN CHIRAL PERTURBATION THEORY}

\author{Cristina Manuel}

\address{Theory Division, CERN \\
CH-1211, Geneva 23, Switzerland \\
(Talk given at the 5th International
Workshop on \\Thermal Field Theories and their Applications, Regensburg 
(Germany), August 1998)}

\maketitle

\begin{abstract}
$\!\!$It is shown how the hard thermal loop approximation can be used
in chiral perturbation theory to study some thermal properties of Goldstone
bosons.
Hard thermal effects are first studied  in the
non-linear sigma model. Then those results are used to obtain
the thermal corrections to the transverse and longitudinal gauge 
field masses in the electroweak theory in the limit of a strongly
interacting Higgs boson. 
\end{abstract}

\baselineskip=15pt
\pagestyle{plain}

\section{INTRODUCTION}
\label{Intro}

This talk is devoted to give a brief account on the emergence of
hard thermal loops \cite{BP}
in the framework of chiral perturbation
theory ($\chi$PT) \cite{We,GassLeut,books}.
First, the one-loop thermal effects arising in
the non-linear sigma model will be discussed \cite{PT,cm1}.
 Second, it will be
shown how to use the previous results in the electroweak model
in the limit where the Higgs boson becomes strongly interacting \cite{cm2}.

The goal of this talk is showing how at a very ``cheap" price
one can obtain the one-loop thermal effective action for soft 
fields in the two above mentioned theories. 
In order to do so one only needs to undertand the symmetry
principles which lie behind hard thermal loops, and how they
were discovered in the context of the high temperature $T$ 
phase of QCD.

Let us first review the  high T phase of QCD and part of the progress
the community achieved in the last years \cite{LeBellac}. 
 The motivation which lead to
the discovery of HTL's was the failure of the naive one-loop
perturbative analysis
in that regime of the theory.
This problem was solved by the now classical works of Braaten and
Pisarski \cite{BP}, with also important contribution of other groups. As we learned
from those works, the naive one-loop computations at finite $T$ are not
complete, since there are one-loop Feynman diagrams, the {\it hard
thermal loops} (HTL's), which are as important as the tree amplitudes,
and therefore they have to be included consistently in all contributions
to non-trivial order in the gauge coupling constant.

Since the pioneering work on HTL's, it is much what we have learned 
about them: their interesting symmetry properties \cite{gaugeinv},
the construction of
effective actions $\Gamma_{HTL} [A]$ from different approaches
\cite{cHTL}, and
the success,  as well as limitations, of the  resummation techniques
\cite{Kobes}.

In this talk it will be shown how the same HTL's
give account of thermal properties of Goldstone bosons.

\section{QCD AT LOW ENERGIES AND TEMPERATURES}

If quarks were massless then the QCD Lagrangian would have
an exact global symmetry $SU_R(N_f) \times  SU_L(N_f)$, where
$N_f$ is the number of quark flavors \footnote{Actually, the global symmetry
is larger, but it is broken by quantum effects that will not be discussed
here.}.
The QCD spectrum of particles
indicates that this global symmetry is spontaneously broken down
to $SU_{R+L}(N_f)$. Associated to the spontaneous breaking of
chiral symmetry there are $N_f ^2 -1$ Goldstone bosons.
For $N_f =3$ those are the octet of pseudoscalar mesons $(\pi's, K's, \eta)$.
The above picture is only approximately valid, since quarks have a
non-vanishing mass, and thus 
the  (pseudo) Goldstone bosons are not massless.

It is possible to study the low energy physics of QCD in the 
framework of chiral perturbation theory ($\chi$PT)
\cite{We,GassLeut,books}. At low energies
the physics of the strong interactions is dominated by the lightest
particles of the QCD spectrum, the (pseudo) Goldstone bosons.

In this talk we will mainly be concerned in studying thermal effects in $\chi$PT
\cite{GLT}
in the chiral limit, that is, in the limit of massless quarks.
This will allow us to understand the properties of a thermal
gas of Goldstone bosons.
The validity of this analysis will be restricted to low $T$.
The reason for this being so is that at low $T$ the strong interactions
should be dominated by the lightest particles, that is,
the  Goldstone bosons. At higher $T$ the contribution in the partition function 
of heavier particles of the QCD spectrum would start to become relevant.

A chiral Lagrangian is expanded in derivatives of the Goldstone fields
and also in the explicit chiral symmetry breaking
parameters, such as the masses of the
three light quarks. The perturbative series in $\chi$PT is not written
in terms of a coupling constant, but rather on 
the energies and masses of the pseudoscalar mesons.

The lowest order chiral Lagrangian is \cite{GassLeut} 

\begin{equation}
{\cal L}_2  = \frac{f_\pi^2}{4}\left( Tr \left(\nabla_\mu \Sigma^{\dagger} \nabla_\mu 
\Sigma\right) + Tr \left( \chi^{\dagger} \Sigma + \chi  \Sigma^{\dagger} \right)\right)
 \ , 
\label{lagnonsiga}
\end{equation}
where $\Sigma$ is a $SU(N_f)$ matrix, which is written in term of the
pseudoscalar mesons $\phi$ as $\Sigma = \exp{(i \phi/f_\pi)}$, and  $f_\pi= 92.4$
MeV
can be identified, to first order,  with the pion decay constant.
 The covariant derivative is defined as
\begin{equation}
\nabla_\mu \Sigma = \partial_\mu \Sigma - i (v_\mu + a_\mu) \Sigma + 
i  \Sigma (v_\mu - a_\mu) \ ,
\end{equation}
$v_\mu$ and $a_\mu$ being external vector and axial vector 
sources, respectively. The field 
$\chi = B (s+i p)$, where $B$ is related to the quark condensate
and $s$ and $p$ are scalar and pseudoscalar external sources,
respectively. 

For the time being all the external sources will be 
taken as $v_\mu = a_\mu = s =p =0$. The generalization of the present
analysis in the presence of external background sources is rather
straightforward, as we will see later on.

In the absence of external sources the Lagrangian (\ref{lagnonsiga}) reduces 
to that of a non-linear sigma model.
The non-linear sigma model has a 
global $SU_R (N_f) \times SU_L (N_f)$ symmetry, where  the field  $\Sigma$
transform as  $\Sigma' (x)  =  U_R \Sigma (x) U_L  ^{\dagger}$, and
$U_{R,L} \in  SU_{R, L} (N_f)$.

To obtain the one-loop effective  action the background field method (BFM)
will be used. The BFM is
a standard technique to evaluate the loop effects generated by a
Lagrangian and  consists in expanding it around the 
solution of the classical equations of motion. 
The one-loop effective action is then obtained after integrating out the
quantum fluctuations.  

In our case one defines   
\begin{equation}
\label{splitS}
\Sigma (x) = \xi(x) h(x) \xi(x)  \ ,
\end{equation}
where  $\bar{\Sigma} = \xi^2$ is the classical solution to the equations of motion
and $h$ is the quantum field.
At this point one  writes $h=\exp{(i {\tilde \phi}/f_\pi)}$, where ${\tilde \phi}$ is a traceless
and
hermitian matrix, 
 and expands the exponentials, keeping only terms up and including 
the quadratic in ${\tilde \phi}$ in the Lagrangian ${\cal L}_2 = {\cal L}_2 ^{(0)}
+ {\cal L}_2 ^{(2)} + \cdots$.
It is possible to express the above terms  as \cite{GassLeut}
\begin{eqnarray}
& {\cal L}_2 ^{(0)} & =  -  f^2_\pi \, Tr (\Delta_\mu)^2 \\
\label{anlag}
& {\cal L}_2 ^{(2)}  & =   \frac{1}{4} Tr (d_\mu {\tilde \phi})^2 
  -  \frac{1}{4} Tr ([\Delta_\mu, {\tilde \phi}])^2 \ ,
\end{eqnarray}
where 
\begin{mathletters}
\begin{eqnarray}
d_\mu {\tilde \phi}& = &  \partial_\mu {\tilde \phi} +  [ \Gamma_\mu, {\tilde \phi}] \ , \\
\Gamma_\mu & = & \frac 12 \left( \xi^\dagger \partial_\mu  \xi + \xi \partial_\mu 
\xi^\dagger \right) \ , \\
\Delta_\mu & = &  \frac 12 \left( \xi^\dagger \partial_\mu  \xi - \xi  \partial_\mu  
\xi^\dagger \right) \ , 
\end{eqnarray}
\end{mathletters}

The transformation properties of $\xi$ and $h$ under the  global 
$SU_R (N_f) \times SU_L (N_f)$ symmetry are
\begin{mathletters}
\begin{eqnarray}
\xi'(x) & = & U_R \, \xi (x) U^{\dagger} (x) = U(x) \xi(x)   U_L  ^{\dagger}  \ ,
\label{compenga1} \\
h'(x) & = & U(x) h(x) U^\dagger (x) \ ,
\label{compenga2}
\end{eqnarray}
\end{mathletters}
$\!\!$where $U$ is a unitary matrix which depends on $\xi(x)$, $U_R$ and
 $U_L  ^{\dagger}$.
The transformation rules obeyed by the new  fields are then deduced from
(\ref{compenga1}) - (\ref{compenga2})
\begin{mathletters}
\begin{eqnarray}
{\tilde \phi}' (x) & = & U (x) {\tilde \phi}(x) U^\dagger (x) \ ,  \\
\Gamma' _\mu (x) & =& U (x) \Gamma _\mu (x) U^\dagger (x) + U (x) \partial_\mu
 U^\dagger (x) \ , \\
\Delta' _\mu (x) & =& U (x) \Delta_\mu (x) U^\dagger (x) \ .
\end{eqnarray}
\end{mathletters}

The field $\Gamma_\mu$ transforms like a connection, while 
$\Delta_\mu$ and ${\tilde \phi}$
transforms
covariantly. Thus, it can be immediately checked that  under the above symmetry
each term in (\ref{anlag}) remains invariant.
The Lagrangian (\ref{anlag}) looks formally as the the bosonic matter part of the
Lagrangian of a non-Abelian gauge
theory, $\Gamma _\mu$ being the  corresponding vector gauge field, and ${\tilde \phi}$
being the bosonic field. There is also an additional coupling between the
$\Delta_\mu$ and ${\tilde \phi}$ fields,
 but appart from that, things  look
the same as in a non-Abelian gauge theory.
This parallelism with a gauge field theory
 is just formal, since there is not a kinetic term for 
$\Gamma _\mu$, neither for  $\Delta_\mu$, and thus those fields do not propagate.
However, this parallelism will prove to be very useful to find the one-loop
thermal effective action for soft background fields.

The one-loop effective action of the non-linear sigma model is
obtained by integrating out the ${\tilde \phi}$ fields. At zero temperature
it can be done by evaluating the determinant of a differential operator,
since the action is quadratic in the ${\tilde \phi}$ fields\footnote{Note also that the jacobian
of the change of variables is one at one-loop order.}.
We will not consider here the $T=0$ one-loop effective action 
(see Ref. \cite{GassLeut}  for that), and concentrate only on the thermal part. 

At this point it seems very obvious that for external background fields
with soft momenta, that is $\ll T$, the one-loop thermal effective action
generated after integrating out the ${\tilde \phi}$ fields in the first term in
(\ref{anlag}) {\it  is exactly the same} as the one that would emerge in a
real gauge field theory, if $\Gamma _\mu$ were a real background  gauge field.
In this way, it is very easy to understand that also HTL's appear
in this model, and that the effective action $\Gamma_{HTL} [\Gamma _\mu]$ also
arises naturally. There is also a tadpole diagram generated by the last
term of (\ref{anlag}). Finally, one finds \cite{cm1}
\begin{eqnarray}
\label{1lteff}
&&   S_2 + \delta S_{2,T}  =  - f_{\pi}^2 (T) \int d^4 x \, Tr (\Delta^2_\mu (x))
\\
&- &\frac{N T^2}{12}\int \frac{d \Omega_{\bf q}}{4 \pi} \int d^4 x \,d^4 y\,
Tr \left(\Gamma_{\mu \lambda}   \frac{Q^\mu Q_{\nu}}{- (Q \cdot
d)^2}  \Gamma^{\nu \lambda}  \right) \ ,
\nonumber
\end{eqnarray}
where
\begin{equation}
\Gamma_{\mu \nu} = \partial_\mu \Gamma_\nu -  \partial_\nu \Gamma_\mu
+  [\Gamma_\mu, \Gamma_\nu]  \ ,
\end{equation}
and $Q= (i, {\bf q})$ is a null vector $Q^2=0$. The angular
integral in (\ref{4.1}) is done over all directions of the three 
dimensional unit vector ${\bf q}$.  The second term of the r.h.s. of
Eq.  (\ref{1lteff}) it is just $\Gamma_{HTL} [\Gamma _\mu]$ as written 
in Ref. \cite{gaugeinv}.

Hard thermal effects change $f_\pi$ into $f_\pi(T)$ as
\begin{equation}
f_{\pi} (T) = f_\pi \left( 1 - \frac{N}{24} \frac{T^2} {f_\pi ^2} \right)
\end{equation}
which agrees with the result computed in the literature (see Ref. 
\cite{GLT,fT,PT}).

Although the same one-loop thermal effective action appears in
two rather different models, it should be clear that their respectively physical
meanings are completely different. In a gauge field theory, 
$\Gamma_{HTL} [A_\mu]$ is a gauge invariant Debye mass term for the 
chromoelectric fields. In the non-linear sigma model  
$\Gamma_{HTL} [\Gamma_\mu]$ describes thermal scattering among 
Goldstone bosons,  respectful with the symmetries of the model.
This can be checked by
 writing $\xi = \exp{ (i \phi/2 f_\pi)}$, and expanding 
the exponentials, so that
\begin{eqnarray}
\Delta_\mu & = & \frac{i}{2 f_\pi} \partial_\mu \phi + O (\frac{1}{f_\pi ^3})
\\ 
 \Gamma_\mu & = &\frac{1}{8 f_\pi ^2} [\phi, \partial_\mu \phi]
+ O (\frac{1}{f_\pi ^4}) \ .
\end{eqnarray}
When the above expressions are plugged into the effective action
 $\Gamma_{HTL} [\Gamma _\mu]$ one can then read the thermal
corrections to the four point functions, six point functions, etc.

Finally, let us explain how one can generalize the above analysis in 
the presence of external vector $v_\mu$ and axial vector $a_\mu$  currents.
Introducing the combinations
\begin{equation}
 F_\mu ^R = v_\mu + a_\mu \ , \qquad  F_\mu ^L = v_\mu - a_\mu \ ,
\end{equation}
and changing the definitions of the $\Gamma_\mu$ and  $\Delta_\mu$
fields as follows
\begin{mathletters}
\begin{eqnarray}
\Gamma_\mu & = & \frac 12 \left( \xi^\dagger \nabla_\mu ^R \xi + \xi \nabla_\mu ^L 
\xi^\dagger \right) \ , \\
\Delta_\mu & = &  \frac 12 \left( \xi^\dagger \nabla_\mu ^R \xi - \xi \nabla_\mu ^L 
\xi^\dagger \right) \ , \\
\nabla_\mu ^l & = & \partial_\mu - i F_\mu ^l \ , \qquad l = R, L  \ ,
\end{eqnarray}
\end{mathletters}
$\!\!$the same one-loop thermal effective action Eq. (\ref{1lteff})
for soft fields is found after integrating out the $\phi$ field.

\section{THE ELECTROWEAK MODEL IN THE STRONGLY INTERACTING
HIGGS BOSON LIMIT}

In the previous Section the one-loop thermal effective action in a 
theory describing the interactions of soft Goldstone bosons has been
computed.  In this Section we will consider a theory where the 
Goldstone bosons are eaten by gauge fields to become massive: we will 
consider the $SU_W (2) \times U_Y(1)$ electroweak model.

The bosonic sector of the electroweak model can be written as \cite{Appel}

\begin{eqnarray}
\label{2.6}
{\cal L} & = &- \frac 12 \,{\rm Tr} (W_{\mu \nu} W^{\mu \nu} ) - 
\frac 14 \,B_{\mu \nu} B^{\mu \nu} \\
& + & \frac 14 \,{\rm Tr} (D_\mu M  D^\mu M ^\dagger)
 -  \frac{\lambda}{4} 
\left(\frac 12 \, {\rm Tr} (M M^\dagger) - \frac{\mu^2}{\lambda} \right)^2 \ ,
\nonumber
\end{eqnarray}
The covariant derivative acting on the matrix $M$ is
\begin{equation}
\label{der}
D_\mu M = \partial_\mu M + i g \, W_\mu M - i \frac {g'}{2}\, M \,B_\mu \tau^3 \ .
\end{equation}

The matrix $M$ is  written in terms of the physical Higgs field $H$
and the would-be Goldstone bosons $\phi^a$ as
\begin{equation}
\label{2.8}
M (x) = \left(v + H(x) \right) \Sigma (x) \ , \qquad 
\Sigma (x) = \exp{(i \frac{{\vec \phi} \cdot {\vec \tau}}{v})}
\end{equation}
where $v= \sqrt{\mu^2 /\lambda}$ is the vacuum expectation value.

This non-linear representation of the Higgs sector
is very suited to study the model in the strongly
interacting limit  $\lambda \rightarrow \infty$. In that 
limit the Higgs mass, $M_H = \sqrt{2 \lambda v^2}$, becomes large, and
the Higgs field can be integrated out. Then
the effective Lagrangian of the electroweak theory reduces 
at tree level to \cite{Appel}
\begin{eqnarray}
\label{2.10}
{\cal L}_{eff} &  = & - \frac 12 \,{\rm Tr} (W_{\mu \nu} W^{\mu \nu} ) - 
\frac 14\, B_{\mu \nu} B^{\mu \nu}  \\
& + &
 \frac {v^2}{4} {\rm Tr} \, (D_\mu \Sigma\, D^\mu \Sigma ^\dagger)\ .
\nonumber
\end{eqnarray}
That is,  the low energy effective theory for the
bosonic sector of the electroweak model
is a gauged $N=2$ non-linear sigma model.

In the unitary gauge, that is, in the gauge where
all the would-be Goldstone bosons are eaten by the gauge fields
(i.e. where $\Sigma =1$), one can read off
the masses of the physical gauge fields  from Eq.  (\ref{2.10}).
The fields $W_\mu ^+$, $Z_\mu$ and $A_\mu$
are defined as
\begin{mathletters}
\begin{eqnarray}
\label{2.11}
W_\mu ^{\mp} & = & \frac {1}{\sqrt{2}} \left(W_\mu ^1 \pm i W_\mu ^2 \right) \ , \\
\label{2.12}
Z_\mu & = & \cos{\theta_W} W_\mu ^3 - \sin{\theta_W} B_\mu \ , \\
A_\mu & = & \sin{\theta_W} W_\mu ^3 + \cos{\theta_W} B_\mu \ ,
\label{2.13}
\end{eqnarray}
\end{mathletters}
$\!\!$where $\theta_W$ is the Weinberg angle, $\tan{\theta_W}= g'/g$. The masses of those
fields are
\begin{equation}
\label{2.14}
M_W = \frac{v g}{2} \ , \qquad M_Z = \frac{v}{2} \sqrt{g^2 + g'^2} \ , 
\end{equation}
while the mass of the photon is $ M_\gamma = 0$.
Recall that the electric charge is defined as
\begin{equation}
\label{2.16}
e = \frac{g g'}{\sqrt{g^2 + g'^2}} \ .
\end{equation}

The one-loop effective action associated to the Lagrangian
(\ref{2.10}) can be computed 
with the help of the background field method.  The gauge fields
are split into background and quantum gauge fields as follows
\begin{mathletters}
\begin{eqnarray}
\label{3.1}
W_\mu  (x) &  = & {\bar W_\mu}  (x) + w_\mu (x)  \ ,  \\
 B_\mu  (x) & = & {\bar B_\mu} 
(x)  + b_\mu  (x)  \ .
\end{eqnarray}
\end{mathletters}
The background gauge fields have been represented by upper
case  letters with a bar,
while the quantum ones are denoted by lower case letters. 
The matrix $\Sigma$ is  split multiplicatively,  exactly as it was done in
Eq.  (\ref{splitS}).

After the splitting of fields is done, the  Lagrangian (\ref{2.10}) is 
separately invariant under background and quantum gauge transformations.

In the spirit of the BFM 
the Lagrangian ${\cal L}_{eff}$ is expanded around the classical
fields, keeping terms which are quadratic in the quantum fluctuations.
To derive the one-loop effective action one has to integrate out the
quantum fields, adding before the
corresponding quantum gauge-fixing and  quantum Faddeev-Popov terms.

In the BFM  it is possible to fix the 
background and quantum gauges independently. In our case, 
and to simplify the computations, it is convenient to choose the
unitary gauge for the background  fields. Then the background 
Goldstone fields disappear completely from the Lagrangian, 
since they are eaten by the background gauge fields to become 
massive. In order to do so it is convenient to use the 
Stueckelberg formalism \cite{Stu}. 
 
If one performs the Stueckelberg transformation
\begin{mathletters}
\begin{eqnarray}
\label{3.17}
& {\bar W}'_\mu   =   \xi^\dagger  {\bar W}_\mu \xi - \frac ig \,\xi^\dagger
\partial_\mu \xi \ , & \\
&{\bar B}'_\mu \frac {\tau^3}{2}  =   \xi ({\bar B}_\mu \frac {\tau^3}{2} )\xi^\dagger
- \frac {i}{g'}\, \xi \partial_\mu \xi^\dagger \ , &  \\
\label{3.18}
& w'_\mu  = \xi^\dagger  w_\mu \xi \ ,  \qquad
b'_\mu  \frac {\tau^3}{2}  =  \xi (b_\mu  \frac {\tau^3}{2}) \xi^\dagger \ ,  & 
\end{eqnarray}
\end{mathletters}
$\!\!$and one writes the Lagrangian in terms of the primed fields, all the
background  fields $\xi$ disappear completely!

The Stueckelberg transformation simplifies drastically the one-loop computations.
Once the computation is finished, the Stueckelberg transformation 
has to be inverted to recover the presence of the background Goldstone
bosons in the final one-loop effective action.

In order to simplify the notation from now on  I will omit the primes in the fields,
keeping in mind that  the transformation has
to be inverted at the end of the computation.

To integrate out the quantum fields a quantum gauge fixing condition 
invariant under the background gauge transformation has to be given.
In the unitary background gauge
the gauge fixing condition for the quantum fields is chosen as
\begin{eqnarray}
\label{3.20}
{\cal L}_{gf}^{(2)} & = & - \frac{1}{a_w} \, {\rm Tr} \left({\bar D}_W ^\mu w_\mu
- \frac 14 a_w g v \phi \right)^2  \\
& - &   \frac{1}{2 a_b} 
\left(\partial^\mu b_\mu + \frac 12 a_b g' v \phi_3 \right)^2 \ ,
\nonumber
\end{eqnarray}
where $a_w$ and $a_b$ are the gauge fixing parameters.
These gauge fixing terms are chosen such as to cancel the
unwanted pieces $\partial^\mu b_\mu \phi_3$ and 
${\rm Tr} (\partial^\mu w_\mu \phi)$ in  ${\cal L}_{eff}^{(2)}$. The form
of the gauge fixing term in an arbitrary background gauge can be 
obtained by inverting the Stueckelberg transformation.

The Faddeev-Popov terms associated to  the gauge fixing (\ref{3.20}) are 
computed as usual. Finally, the complete one-loop quantum Lagrangian 
reads in Minkowski space
\begin{eqnarray} 
\label{3.21}
&& {\cal L}^{(2)}_{eff} + {\cal L}_{gf}^{(2)} +{\cal L}_{FP}^{(2)} \\
  & = &  {\rm Tr}\left(  w_\mu (g^{\mu \nu} {\bar D}^2 _W + 
\frac{1 - a_w}{a_w} {\bar D}_W^\mu {\bar D}_W^\nu + 2 i g {\bar W}^{\mu \nu} )
 w_\nu 
\right) \nonumber   
\\
& + & \frac 12 b_\mu \left( g^{\mu \nu} \partial^2 + \frac{1 - a_b}{a_b}
\partial^\mu \partial^\nu \right) b_\nu  \nonumber \\
&  + & M^2_W \, {\rm Tr}  (  w_\mu   w^\mu) + \frac {M^2_B}{2}\, b_\mu b^\mu - g g'
v^2 w_\mu ^3
b^\mu  \nonumber \\
&+ & \frac 14 \, {\rm Tr} ({\bar d}_\mu \phi)^2 - \frac 14 \, {\rm Tr} [{\bar \Delta}_\mu, \phi]^2 
- \frac{a_w M^2_W}{4} {\rm Tr} \phi^2 - \frac{a_b M^2_B}{2} \phi^2_3 \nonumber \\
&+ &  2 v\, {\rm Tr} \left( (g w_\mu - g' b_\mu \frac{\tau^3}{2}) {\bar \Gamma}^\mu \phi \right) \nonumber \\
& - & \eta^{\dagger}_a \left(\delta^{a b}  {\bar D}^2 _W + \delta^{a b} a_w M^2_W \right) \eta_b
 \nonumber 
\end{eqnarray}
where $M^2_B = g'^2 v^2/4$ and
\begin{mathletters}
\begin{eqnarray}
\label{3.22}
{\bar D}_W^\mu & = & \partial^\mu + ig [ {\bar W}^\mu, \, ] \ , \\
\label{3.23}
{\bar d}_\mu \phi & = & \partial_\mu \phi + [ {\bar \Gamma}_\mu, \phi] \ , \\
\label{3.24}
{\bar \Gamma}_\mu & = & \frac i2 \left( g {\bar W}_\mu + g' {\bar B}_\mu
\frac{\tau^3}{2} \right) \ , \\
\label{3.25}
{\bar \Delta}_\mu & = & \frac i2 \left( g {\bar W}_\mu - g' {\bar B}_\mu
\frac{\tau^3}{2} \right) \ . 
\end{eqnarray}
\end{mathletters}

The ghost fields $\eta_a$ are 
associated to the $w_\mu ^a$ quantum
fields. Since the ghost associated to the $b_\mu$ field does not couple
to any background external field, it has been omitted in Eq. (\ref{3.21}).

The one-loop Lagrangian (\ref{3.21}) is written in the unitary background
gauge. It can be obtained in an arbitrary background gauge  by
inverting the Stueckelberg transformation. However, it is much simpler
to integrate out the quantum fields first, and invert the transformation
afterwards to obtain the one-loop effective action in a general 
background gauge.  

For soft background fields the leading thermal corrections arise when
the internal quantum fields are hard \cite{BP}, that is, of energy $\sim T$. If one
neglects corrections of order $M_W /T$ and $M_Z /T$ in the final answers,
then it is possible to neglect those masses for the quantum fields. 
In other words, for hard quantum fields the
terms $\partial^2$ of the Lagrangian are of the order $T^2$, which are dominant as compared
to the terms $M^2_{W,B}$, which therefore will be neglected.

The computation simplifies once the masses of the quantum fields are neglected. 
One encounters here the same one-loop thermal amplitudes,
the HTL's, which appear in the BFM of Yang-Mills theories
\cite{BP}, as well as in the non-linear sigma model in the presence
of external background sources \cite{cm1}. There are also new types of vertices 
in (\ref{3.21}), which do not appear in the BFM studies of the  previous mentioned theories:
those which couple quantum gauge fields and quantum Goldstone bosons.
However, a power counting analysis shows that the one-loop 
thermal corrections generated by those vertices
are subleading as compared to the HTL's, and therefore they will be
neglected.

The one-loop thermal effective action for soft background gauge fields
is then a combination of the one which appears in a Yang-Mills theory
and the one in the non-linear sigma model in the presence of external sources.
By translating those results to our case one  finds the following
one-loop thermal effective action \cite{cm2}

\begin{eqnarray}
\label{4.1}
&& S_{eff} + \delta S_{eff,T}  = \\
&&  \int d^4 x \left\{
- \frac 12 \,{\rm Tr} ({\bar W}_{\mu \nu} {\bar W}^{\mu \nu} )  \right.- 
\frac 14 \, {\bar B}_{\mu \nu} {\bar B}^{\mu \nu} \nonumber \\
&&   \left. + \frac {v^2(T)}{4} 
\,{\rm Tr}\left( g {\bar W}^\mu - g' {\bar B}^\mu \frac{\tau^3}{2} \right)^2 \right\}
\nonumber  \\
&- &\frac{T^2}{6}\int \frac{d \Omega_{\bf q}}{4 \pi} \int d^4 x \,d^4 y\,
{\rm Tr} \left({\bar \Gamma}_{\mu \lambda}   \frac{Q^\mu Q_{\nu}}{- (Q \cdot
{\bar d})^2}  {\bar \Gamma}^{\nu \lambda}  \right) \nonumber \\
& + &  \frac{g^2 T^2}{3}\int \frac{d \Omega_{\bf q}}{4 \pi} \int d^4 x \,d^4 y\,
{\rm Tr} \left({\bar W}_{\mu \lambda}   \frac{Q^\mu Q_\nu}{- (Q \cdot
{\bar D}_W)^2}  {\bar W}^{\nu \lambda}  \right)
\nonumber
\end{eqnarray}
where ${\bar W}_{\mu \nu}$,  ${\bar B}_{\mu \nu}$ are the field strengths
of the corresponding background gauge fields, and
\begin{eqnarray}
\label{4.2}
v(T) & =  & v \left( 1 - \frac{1}{12} \frac {T^2}{v^2} \right) \ , \\
\label{4.3}
{\bar \Gamma}_{\mu \nu} & = & \partial_\mu {\bar \Gamma}_\nu
- \partial_\nu {\bar \Gamma}_\mu +
 [{\bar \Gamma}_\mu,{\bar \Gamma}_\nu] \ , 
\end{eqnarray}

Let us remind the meaning of each term of Eq. 
(\ref{4.1}). The two first terms are
the kinetic pieces for the soft background gauge fields.
The last piece in Eq. (\ref{4.1})
is the HTL effective action for the non-Abelian gauge field ${\bar W}_\mu$,
and it is generated by considering the one-loop thermal effects of the hard
quantum gauge field $w_\mu^a$, and the quantum ghosts $\eta^a$ \cite{BP}.
The third and fourth terms in Eq. (\ref{4.1})
arise after considering the one-loop thermal effects of the hard quantum Goldstone bosons
$\phi^a$,
(see the previous Section
 with the following identifications: $F^R_\mu = -g {\bar W}_\mu$,
$F^L_\mu = -g' {\bar B}_\mu \frac{\tau^3}{2}$, and $\xi=\xi^\dagger=1$.).

In the static limit the non-local terms of Eq. (\ref{4.1}) become local, and
then appart from the kinetic terms for the gauge fields, one has
\begin{eqnarray}
\label{5.1}
\delta {\cal L}_{eff,T}^{static} & = &
\frac {v^2(T)}{4} 
\,{\rm Tr}\left( g {\bar W}^\mu - g' {\bar B}^\mu \frac{\tau^3}{2} \right)^2 \\
& +& \frac{T^2}{12} \,{\rm Tr}\left( g {\bar W}_0 + g' {\bar B}_0 \frac{\tau^3}{2} \right)^2
+  \frac{2 g ^2T^2}{3} {\rm Tr}({\bar W}_0)^2 \ . \nonumber
\end{eqnarray}

If one expresses Eq. (\ref{5.1}) in terms of the physical fields ${\bar W}^+$, ${\bar Z}_\mu$
and ${\bar A}_\mu$,  one obtains the corrections to their longitudinal and
transverse masses,  plus couplings of the ${\bar Z}^0 $ 
with ${\bar A}^0$ fileds.

The longitudinal and transverse gauge modes get different thermal corrections to their masses.
The thermal masses for the transverse modes are
\begin{eqnarray}
\label{4.x}
 M^2_{W,t} (T) & = &\frac{g^2 v^2 (T)}{4} \ , \\
 M^2_{Z,t}(T) & = &\frac{(g^2 + g'^2) v^2 (T)}{4} \ , \\
 M^2_{\gamma, t} (T) & = & 0  \ ,
\end{eqnarray}
while  for the longitudinal ones are
\begin{eqnarray}
\label{4.xx}
 M^2_{W,l} (T) & = & \frac{g^2 v^2 (T)}{4} + \frac{3 g^2 T^2}{4}
\ ,  \\
  M^2_{Z,l}(T) & =  &\frac{(g^2 + g'^2) v^2 (T)}{4}  
 + 
\frac{  (g^2+g'^2) T^2}{12}  \\
& \times & \left( \cos^2{\theta_W} 
 -  \sin^2{\theta_W}\right)^2 
+ \frac{2  g^2 T^2}{3} \cos^2{\theta_W}  \ , \nonumber
 \\
 M^2_{\gamma, l} (T) & = & e^2 T^2  \ .
\end{eqnarray}
To express the  electric thermal mass of the photon in terms of the electric charge $e$, use of the relation 
(\ref{2.16}) has been made.  
The above results agree were first obtained in Ref.
\cite{Gavela}.

\section{CONCLUSIONS}

It has been shown how in the chiral limit hard thermal loops 
appear in the framework of $\chi$PT. This fact allows us to
study certain thermal properties of Goldstone bosons with ease,
both in a theory where those are real particles, or where those
are unphysical since they are eaten by the gauge fields to become
massive.

Some other applications of the HTL's techniques in  $\chi$PT have already
been exploited. In the literature HTL's have been used to compute thermal
corrections to the anomalous decay $\pi^0 \rightarrow \gamma \gamma$
\cite{anom},
to the Wess-Zumino-Witten action \cite{cm1}, or to the electromagnetic mass
difference of pions\cite{CMNR}. Further applications of HTL's in $\chi$PT will
surely be found.

{\bf Acknowledgements:} 
I would like to thank the organizers of the workshop for financial
support.

\end{document}